\documentclass[aps,prb,twocolumn,showpacs]{revtex4}
\usepackage{graphicx}

\begin{document}
\title{Two $s$-wave solutions for superconductivity in the extended Hubbard model}
\author{Maciej Bak}
\email[]{karen@delta.amu.edu.pl} \affiliation{Institute of
Physics, A. Mickiewicz University, Umultowska 85, 61-614 Pozna\'n,
Poland}

\begin{abstract}
The existence  of more than one solution for s-wave pairing in the
extended Hubbard model is not often realized. This possibility was
noted by Friedberg et al. [Phys. Rev. B50, 10190 (1994)] in the
case of two electrons on a lattice, without further analysis. In
the present paper, the second solution is found also in the case
of superconductivity of extended s-wave symmetry in the extended
Hubbard model. The properties of both s-wave solutions are
examined by mean-field methods and thresholds for their appearance
are given. A possibility of the first-order transition is
discovered and modifications of the phase diagram are calculated.
Results of this paper in the limit of low electron density are
also applicable to the bound, two-electron pairs.
\end{abstract}


\maketitle

\section{Introduction}
The discovery of high-temperature superconductivity (HTS) has
triggered an intensive effort to reveal its mechanisms. This aim
has not been attained yet but the amount of knowledge concerning
HTS is substantial. From this wealth of data certain key issues
emerge, which must be addressed by the eventual microscopic
theory. One of them is question of the symmetry of the order
parameter. Instead of originally accepted $s$-wave, nowadays it is
believed that the hole-doped HTS materials are predominantly of
the $d_{x^2-y^2}$-wave symmetry.~\cite{orenstein,harlingen}
Nevertheless the research concerning the $s$-wave symmetry is not
to be abandoned. The $s$-wave symmetry is found in
fullerides,~\cite{capone} perovskites~\cite{lin} and
$MgB_2$,~\cite{seneor} not from the HTS group but very interesting
materials though. The order parameter of $s$ symmetry can appear
as subdominant one in $d$-wave superconductors due to orthorombic
distortions of the crystal.~\cite{li,donovan,kouznetsov} There is
evidence of the $s$ type HTS in electron-doped
materials~\cite{alff,skinta,biswas,skinta2} and possibility of the
$s$ to $d$ change of the symmetry of the order parameter with
increasing doping in the hole-doped materials.~\cite{micnas,yeh}
Another scenario with $s$-wave symmetry includes the possibility
of different order parameters in the bulk and on the surface of
HTS superconductor.~\cite{zhang} As there are reports of different
mechanisms responsible for pairing and phase coherence there is
possibility of pseudogap having $s$ symmetry, different from the
$d$ symmetry of the order parameter.~\cite{deutscher,mourachkine}
Of other possibilities we can not exclude possibility of finding
an $s$-wave symmetry pairing in heavy fermions, which are
notorious for multiplicity of phases.

There are few rigorous results concerning $s$-wave
superconductivity or BCS in general. We know that in the dilute
limit BCS mean-field equations go over to the Schr\"{o}dinger
equation for the bound two-electron pair,~\cite{micnas,nozieres}
i.e. in this limit the BCS equations become {\em exact}. Another
result was given by Randeria {\em et al.}~\cite{randeria} stating
that creating two-electron bound state of $s$-wave symmetry is a
necessary and sufficient condition for BCS-type superconductivity
to exist in two dimensions in a dilute limit. The results of
applying mean-field approximations to the Hubbard-type
Hamiltonians as well as the properties of the bound pairs are well
known (e.g. Refs
~\cite{micnas,pistolesi,bakmic,doktorat,robaszkiewicz,blaer,friedberg}).
One of the seldom noticed aspects of mean-field treatment of the
extended $s$-wave pairing and bound pairs of this symmetry is
subject of this paper.

\section{Formalism}
Let's consider extended Hubbard model in the standard notation:
\begin{equation}
    H=\sum_{ij\sigma}(t_{ij}-\mu\delta_{ij})c^\dagger_{i\sigma}c_{j\sigma}+
        U\sum_i n_{i\uparrow}n_{i\downarrow}+
        W\sum_{ij\sigma\sigma'}n_{i\sigma}n_{j\sigma'}
\end{equation}
where we sum over nearest neighbors (n.n.), $U$ and $W$ are on-
and intersite interactions respectively, $t_{ij}$ is
nearest-neighbor hopping integral and $\mu$ - chemical potential.
In the mean-field approach the above Hamiltonian takes the
form~\cite{micnasRRT}:
\begin{equation}
    H=\sum_{k\sigma}(\varepsilon_k-\overline\mu\,)\,n_{k\sigma} -
        \sum_k (\Delta_k
        c^\dagger_{k\uparrow}c^\dagger_{-k\downarrow}+H.c.)+const.
\end{equation}
where, using order parameter $<c_{-q\downarrow}c_{q\uparrow}>$, we
introduced a gap $\Delta_k$:
\begin{eqnarray}
    \Delta_k={1\over N}\sum_q
        V_{kq}<c_{-q\downarrow}c_{q\uparrow}>\\
    V_{kq}=-U-W\gamma_{k-q}\\
    \gamma_k =\sum_{\delta}^z\exp(i{\bf k}\cdot\mbox{\boldmath $\delta$})\\
    \varepsilon_k=-t\gamma_k\\
    \overline\mu=\mu-(U/2+W z)n
\end{eqnarray}
$z$ is coordination number; for the full discussion of the model
the reader is referred to the Ref.~\cite{micnasRRT} It is easy to
derive the self-consistent equation for the energy gap:
\begin{equation}
    \Delta_k={1\over N}\sum_q V_{kq}\frac{\Delta_q}{2E_q}\tanh{\beta
    E_q/2}
\end{equation}
where the quasiparticle energy:
\begin{equation}
    E_q=\sqrt{(\varepsilon_q-\overline\mu)^2+|\Delta_q|^2}
\end{equation}
and $\beta=1/(k_B T)$ where $T$ is temperature and $k_B$ Boltzmann
constant. To solve the model we usually make the
ansatz:~\cite{micnas}
\begin{equation}\label{ansatz}
    \Delta_k=\Delta_0+\Delta_{\gamma} \gamma_k+\Delta_{\eta}\eta_k
\end{equation}
where $\eta_k=2(\cos k_x - \cos k_y)$ and particular terms refer
to on-site $s$-, extended $s$- and $d$-wave pairings.
Self-consistent equations for the gap separate into the $s$-wave
and $d$-wave part.

Finally what is left for solving for general, non-zero $U$ and $W$
in the ground state is a set of three self-consistent equations
for $\Delta_0$, $\Delta_{\gamma}$ and chemical potential
$\overline\mu$:
\begin{eqnarray}\label{delta0}
    \Delta_0=-U {1\over N}
        \sum_q(\Delta_0+\gamma_q\Delta_{\gamma})\frac{1}{2E_q}\\\label{delta1}
    \Delta_{\gamma}=-{W\over z} {1\over N}
        \sum_q\gamma_q(\Delta_0+\gamma_q\Delta_{\gamma})\frac{1}{2E_q}\\\label{n}
    n-1=-{1\over N}\sum_q\frac{(\varepsilon_q-\overline\mu)}{E_q}
\end{eqnarray}
in the case of the extended $s$-wave pairing, and
\begin{equation}
    1=-{W\over z} {1\over N}\sum_q\eta_q^2{1\over 2E_q}
\end{equation}
in the case of the $d$-wave pairing. In the following we will
focus on the $s$-wave pairing only. The results described in the
remainder of the paper are calculated for the rectangular density
of states (DOS) with $z=4$, so they approximate two dimensional
square lattice. For the reasons that will be explained later on
the author supposes that they are of more general nature.

\section{Results}
\begin{figure}
\includegraphics[scale=0.5]{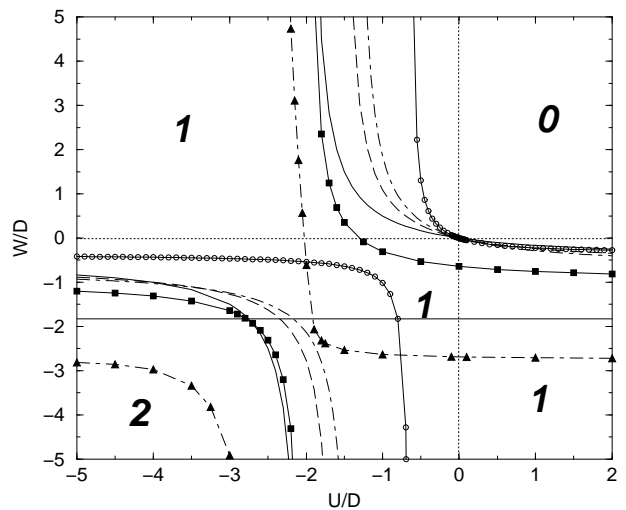}
\caption{\label{UWphDg} Critical values for existence of extended
$s$-wave superconductivity calculated for the rectangular DOS, for
electron densities $n=0$ (also $n=2$, full line) and $n=0.2$ (also
$n=1.8$, dashed line). Dot-dashed and full line with circles are
thresholds for $n=0.2$, calculated for the exact DOS in two
dimensions (2d) and in three dimensions (3d, simple cubic lattice)
respectively. Exact 2d threshold for $n=0$ is the same as for
rectangular DOS. Exact 3d threshold for $n=0$ is given by solid
line with squares. Dotted lines are the axes of coordination
system. Dot-dashed lines with triangles are crossover boundaries
for $n=0.2$ and exact 2d DOS. Crossover boundaries for $n=0$
coincide with the threshold lines in 2d. Full, flat, horizontal
line is a threshold for the $d$-wave pairing, calculated for
$n=0$. Numbers "0", "1" and "2" are the names of the areas
described in the text. $U$ and $W$ are in the half-bandwidth units
$D=4t$ in 2d and $D=6t$ in 3d.}
\end{figure}

Equations~(\ref{delta0})-(\ref{n}) in the dilute limit yield
Schr\"{o}dinger equation for a bound pair, as was told in the
Introduction. As is well known, in two dimensions infinitesimally
small on-site attraction creates bound pair for $W=0$. For $W>0$ a
threshold must be crossed to obtain a bound state. Analytic
formula for the threshold and $U$-$W$ phase diagram are given for
example in.~\cite{micnas,kornilovitch} The less known facts are as
follows:

(i) The threshold formula for the gap, analogical to the one given
in,~\cite{micnas,kornilovitch} can be derived from BCS-MFA
equations (\ref{delta0})-(\ref{n}) {\em without} resorting to the
limit $n\rightarrow 0$. Using rectangular density of states it
reads:
\begin{equation}\label{wcr}
    \frac{W_{crit}}{4t}=\frac{-1}{1+(n-1)^2\,(1+16t/U)}
\end{equation}
Let's note that the limit $n\rightarrow 0$ yields correct result
of Refs.~\cite{micnas,kornilovitch} The use of the exact density
of states does not bring changes to the value of $W_{crit}$ for
$n\rightarrow 0$ but the changes appear and increase with growing
$n$, as shown in Fig.~\ref{UWphDg}.

(ii) The formula (\ref{wcr}) is valid on the {\em entire} $U$-$W$
plane, both for positive and negative values of $U$ and $W$. The
cited papers show the figure of the fourth quarter of the
coordination system only. The full plot is given in
Fig.~\ref{UWphDg}. As we can see in the third quarter of the
coordination system, for $U$ and $W$ attractive enough, a second
threshold line is obtained! (In the case of two-electron bound
pair second solution was also noted in Ref.~\cite{friedberg} but
without further analysis).

This way the area in Fig.~\ref{UWphDg} can be divided into three
main parts: Area "0" - to the right of the rightmost threshold
line, mostly first quarter ($U>0$, $W>0$) and small parts of the
second and fourth quarter of the coordination system (not
exceeding asymptotic values: $U_{as}/t=-16(n-1)^2/(1+(n-1)^2)$ and
$W_{as}/t=-4/(1+(n-1)^2)$). There are no bound states and no gap
for these parameters, the solutions describe antibonding states.
Area "1" -- between both threshold lines, mostly second and fourth
quarter -- in this area there is only one solution of the
equations (\ref{delta0})-(\ref{n}), which exists even for infinite
on- or intersite repulsion. The further from the rightmost
threshold line we are the stronger is the binding between
electrons in the pairs. Points of the same binding energy form a
line approximately parallel to the rightmost threshold line. Area
"2" -- below the second threshold line, in the third quarter --
for parameters from this area two solutions of equations
(\ref{delta0})-(\ref{n}) exist: one of the type of the "area 1",
with large binding energy, given by a "distance" from the first
threshold line; the other, with smaller binding energy and smaller
gap, given by the "distance" from the second threshold line. We
will call these solutions type 1 and type 2, respectively. Let's
note, that the range of $U$, for which both types of pairing
exist, increase with increasing $n$ (the asymptote moves from
$U_{as}/4t=-2$ for $n=0$ to $U_{as}=0$ for $n=1$) while analogical
range of $W$ decrease (the asymptote moves from $W_{as}/4t=-1/2$
for $n=0$ to $W_{as}/4t=-1$ for $n=1$). The properties of both
types of pairing are described in the following.

\begin{figure}
\includegraphics[scale=0.5]{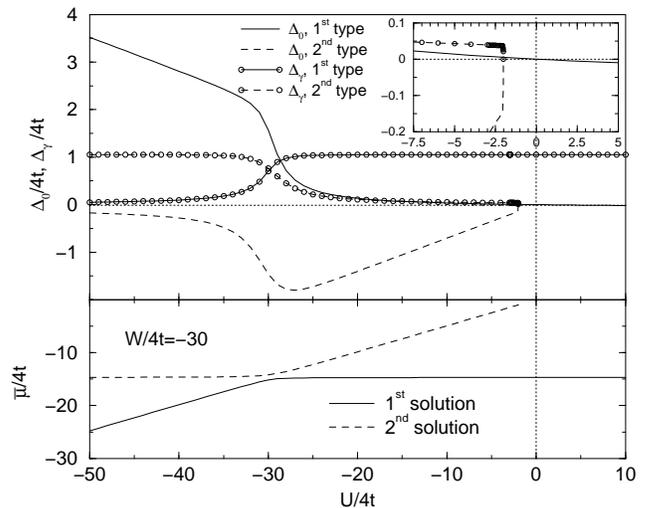}
\caption{\label{vsU} Type 1 (full line, full line with circles)
and type 2 (dashed line, dashed line with circles) solutions of
the extended Hubbard model vs $U/4t$ for $W/4t=-30$ and $n=0.01$.
Upper panel shows $\Delta_0$ and $\Delta_{\gamma}$ and lower panel
$\overline\mu$. Dotted lines show axes of the coordination system.
In the inset the enlargement of the area near the origin of the
coordination system in the upper panel is shown.}
\end{figure}

For fixed $W$, for $U\rightarrow -\infty$, type 1 solution
displays what we will call here "pure-s" type behavior:
$|\Delta_0|\sim U$, $\Delta_{\gamma}\rightarrow 0$,
$\overline\mu\sim U/2$ (if $W>0$ then $\Delta_{\gamma}<0$, if
$W<0$ then $\Delta_{\gamma}>0$). When we move on the $U$ axis from
left to right, towards smaller values of $|U|$ then, for certain
intermediate value of $|U|$ (not too large if $W$ is not large and
attractive) this "pure-s" behavior disappears. If $W>W_{as}$ then
with increasing $U$ we must reach threshold value $U_{crit}$ and
for $U\rightarrow U_{crit}$ $\Delta_0$ decreases non-linearly to
0, $|\Delta_{\gamma}|$ goes through extremum and also decreases to
0 while $\overline\mu$ goes to constant, non-zero value. If
$W<W_{as}$ and solution of type 1 exists in the whole range of $U$
from $-\infty$ to $+\infty$, then for $U\rightarrow +\infty$  all
three quantities $\Delta_0$, $\Delta_{\gamma}$ and $\overline\mu$
go to the constant values: $\Delta_0$ to a small negative one and
$\Delta_{\gamma}$ to a larger, positive one (see Fig.~\ref{vsU}).
When $W<0$ and $|W|$ is large enough, then "pure-s" behavior
disappears for $U\approx W$ instead of small to intermediate
values of $|U|$, as shown in Fig.~\ref{vsU}.

The behavior of type 2 solution is somewhat complementary: for
$U\rightarrow -\infty$, when type 1 solution shows "pure-s"
behavior, type 2 solution is constant. Moving right on the $U$
axis type 2 solution goes through a transition for the same range
of $U$ as type 1 solution, and from now on it behaves in "pure-s"
way until $U$ reaches the threshold for type 2 solutions, where
$\Delta_0$ and $\Delta_{\gamma}$ disappear and $\overline\mu=$
const. Let's note, that the slope of linear dependence of
$\Delta_0$ vs $U$ differs just by the sign from the analogical
slope of type 1 solution and that $\Delta_{\gamma}$ is of opposite
sign to $\Delta_0$ in type 2 solution.

\begin{figure}
\includegraphics[scale=0.5]{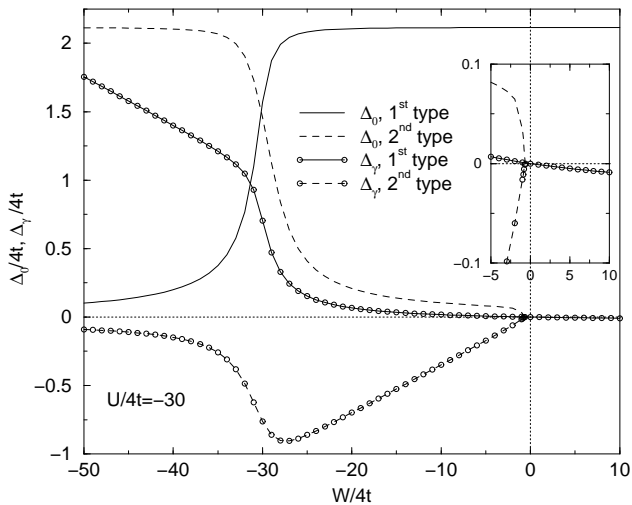}
\caption{\label{vsW} Type 1 (full line, full line with circles)
and type 2 (dashed line, dashed line with circles) solutions of
the extended Hubbard model vs $W/4t$ for $U/4t=-30$ and $n=0.01$.
Dotted lines show axes of the coordination system. The dependence
of $\overline\mu$ on $W$ is indistinguishable in the scale of the
picture from the dependence of $\overline\mu$ on $U$ in
Fig.~\ref{vsU} (apart from different thresholds for type 2
solutions), so it was not shown. In the inset the enlargement of
the area near the origin of the coordination system is shown.}
\end{figure}

The constant values reached by the type 1 solution for
$U\rightarrow +\infty$ are the same as the ones reached by the
type 2 solution for $U\rightarrow -\infty$. These constants for
small $n$ and $|W|$ large enough can be calculated as:
\begin{eqnarray}\label{limit1a}
    \overline\mu=W/2\\
    \Delta_{\gamma}/\Delta_0=-|\overline\mu|\\\label{limit1b}
    \Delta_0\approx -0.6\sqrt{n}
\end{eqnarray}
As shown in Fig.~\ref{vsW} for fixed $U$, when $W$ is varied,
instead of "pure-s" behavior type 1 solution displays "pure
extended s" behavior for $W\rightarrow -\infty$:
$\Delta_{\gamma}\sim |W|$, $\Delta_0\rightarrow 0$,
$\overline\mu\sim W/2$. Apart from that, the behavior of solutions
vs $W$ is analogous to the behavior of the solutions vs $U$: for
$W\rightarrow +\infty$ type 1 solution goes to the constant values
(now $\Delta_{\gamma}$ to the small negative one while $\Delta_0$
to the bigger, positive one) and for $W\rightarrow -\infty$ type 2
solution goes to the same constants. For small $n$ and $|U|$ large
enough these constants are:
\begin{eqnarray}\label{limit2a}
    \overline\mu=U/2\\
    \Delta_0/\Delta_{\gamma}=-4|\overline\mu|\\\label{limit2b}
    \Delta_{\gamma}\approx -\sqrt{n/8}
\end{eqnarray}
Large values of fixed parameters in Figs~\ref{vsU} and \ref{vsW}
were chosen to emphasize the characteristic behavior of the
examined quantities. The plot of chemical potential $\overline\mu$
vs $U$ for $W/4t=-30$ is practically indistinguishable from the
plot of $\overline\mu$ vs $W$ for $U/4t=-30$ in the scale of the
picture (apart from the different critical values for which
$\overline\mu$ in the type 2 solutions appears) so it was not
shown again in Fig.~\ref{vsW}.

A point to notice is that the gap parameter $\Delta_0$ remains
finite even in the limit $U\rightarrow\infty$, despite the fact
that on-site pairing amplitude
$<c^\dagger_{i\uparrow}c^\dagger_{i\downarrow}>$ must vanish in
this limit.~\cite{bastide,rnm}

 Let's note that the considerations above answer
the unspoken question, why the equations for extended $s$-wave
pairing yield formally (and numerically) the same solution in the
two very different limits: infinite attraction and infinite on- or
intersite repulsion. Despite they are numerically the same,
nevertheless they are two different solutions of two different
types of pairing!

In the limit of large $|U|$ or large $|W|$ the solutions for
$\Delta_0$ and $\Delta_{\gamma}$ are symmetric or antisymmetric
regarding to the change $n\rightarrow 2-n$. For large, negative
$U$ (or $W$) and fixed $W$ (or $U$)
 $\Delta_0$ ($\Delta_{\gamma}$) of "pure-s" (of "pure extended s") type 1 solution is
symmetric, reaching maximum for the half-filled band, while
$\Delta_{\gamma}$ ($\Delta_0$) is antisymmetric: positive (or
negative) for less than half-filled band and of opposite sign for
more than half-filled band, crossing through 0 for $n=1$. In the
opposite limit $U$ (or $W$) going to $+\infty$ the behavior of
$\Delta_0$ and $\Delta_{\gamma}$ in "type 1" solutions change to
the reverse: $\Delta_0$ ($\Delta_{\gamma}$) becomes antisymmetric
and $\Delta_{\gamma}$ ($\Delta_0$) symmetric. The behavior vs $n$
of type 2 solutions for $U$ or $W\rightarrow -\infty$ is the same
as behavior of respective "pure-s" or "pure extended s" type 1
solutions in the limit $U$ or $W\rightarrow +\infty$. This is the
correct behavior if the limiting, constant values reached by the
type 1 solutions are to be identical to the type 2 limiting,
constant solutions for every $n$. For smaller values of the
parameters $|U|$, $|W|$ there is an area in parameter space of
"type 2" solution, where both $\Delta_0$ and $\Delta_{\gamma}$ are
symmetric in the band, though of opposite sign.

A peculiarity of solutions is the existence of critical charge
density $n_c$. For a given value of the fixed parameter ($U$ or
$W$) there exists $n_c$ such that for $n$ in the range
$(n_c,2-n_c)$ the transition  between the "pure (extended) s" part
of one solution and the "constant" part of the same solution  is
no longer smooth, but both parts join in an non-continuous way: we
have a first-order transition, where $\Delta$'s and derivative of
$\overline\mu$ jump. Some approximate values of $n_c(U/4t,W/4t)$
for type 1 solution, given as a function of parameters for which
the jump takes place, are: $n_c(-29.1,-30)\approx 0.0475$,
$n_c(-9.1,-10)\approx 0.15$, $n_c(-2.75,-3.75)\approx 0.5$,
$n_c(-1.1,-2)\approx 0.8$. We can see here that for $|W|$ large
enough and attractive the jump takes place for $U\approx W$, as
was stated before. The values of $|U|$ and $|W|$ required to
obtain any given $n_c$ are much larger in type 2 solution than in
type 1 solution.

To complete the  examination of the $n$-dependence let's notice
that at half-filled band solutions fully separate and we obtain
true pure-$s$ solution, with $\Delta_\gamma=0$ and no threshold
for pairing and true pure extended $s$ solution with $\Delta_0=0$
and threshold $W=-4t$ ($W\approx -3.24 t$ for exact 2d density of
states).~\cite{micnasRRT}

Another point to notice is that we can add the third and fourth
solution to the two found, by changing $\Delta_0$ and
$\Delta_{\gamma}$ to $-\Delta_0$ and $-\Delta_{\gamma}$. The
number of nontrivially-different solutions seems to depend on the
number of coupled terms from the ansatz Eq.~(\ref{ansatz}) and
does not depend on the lattice dimensionality. In Fig. 1 there is
a plot of threshold curves for simple cubic, three dimensional
(3d) lattice. Their behavior is analogous to the 2d case. In
particular the threshold line for the type-1 pairing goes through
the point $(U,W)=(0,0)$, even for small electron densities (except
from $n=0$). This does not contradict the fact that for $U=0$ we
have critical value for $|W|$ to bound electron pair in 3d. The
existence of bound pairs is prerequisite for superconductivity in
a dilute limit only in 2d\cite{randeria}. A curve for two-electron
bound pair in 3d can be obtained from
Eqs.~(\ref{delta0})-(\ref{n}), by solving them for $n=0$. This
curve is also given in Fig. 1, plotted in a certain distance from
other 3d curves for $n>0$, in agreement with
Refs.\cite{micnas,kornilovitch}

There appears a question: which of the two solutions will be
physically realized? The type 1 pairing has larger binding energy,
higher critical temperature and lower free energy so this is the
solution to be found even in the area of parameter space where
both solutions exist. In this area care should be taken not to mix
the two solutions especially for $U\approx W$, where the absolute
values of solutions of type 1 become close to the respective
values of solutions of type 2 and also in the limit of large
negative $U$ or $W$, where equations can easily yield type 2
solution numerically.

We must also remember that only the relative stability of two
extended $s$-wave solutions was examined in the above
considerations. The question of stability of another symmetry
pairings, e.g. $d$- or $p$-wave, was not addressed at all. While
$p$-wave pairing can be diminished by adding antiferromagnetic
coupling to the Hamiltonian, $d$-wave can not. From Fig.~1 we can
see that for large enough $|W|$, $d$-wave pairing can surely have
bigger binding energy than second extended $s$-wave solution. As
we know it can win the competition even with the first extended
$s$-wave solution (with the lower free energy -- see
e.g.~\cite{micnas}). In establishing the detailed phase diagram we
must thus take into account the competition of pairings with other
symmetries, including $s+d$ or $s+id$.

Another question is how physical is the first order transition
found, isn't it just an artefact of the MFA, especially that it
appears for larger $n$? This question could be probably answered
by numerical simulations. If this effect is physical it should
have influence on the all superconducting properties of the
superconductor with extended $s$-wave superconductivity. In
particular one can ask whether the found first order transition
could be an argument in favor of non-smooth crossover between BCS
and local pairs limits in the models of extended $s$-wave
superconductivity. It seems unlikely as we notice that $|U|$ and
$|W|$ being the arguments of $n_c$ cited before in the text,
decrease for growing  $n_c$ while the BCS-Bose boundary
(calculated form the Leggett's criterion,~\cite{leggett} i.e.,
that chemical potential crosses the bottom of the band) moves
toward larger values of $|U|$ and $|W|$ with increasing electron
density, so that the two boundaries do not coincide. Crossover
boundaries for $n=0$ and $n=0.2$ are plotted in Fig.~\ref{UWphDg}.
There are two branches of crossover boundary corresponding to the
two threshold lines. For $n=0$ they coincide with the threshold
lines, with increasing $n$ they move left, towards larger and
larger values of $|U|$ and $|W|$. Again, if other than $s$-wave
symmetry pairing is realized, this problem is irrelevant.

\begin{figure}
\includegraphics[scale=0.51]{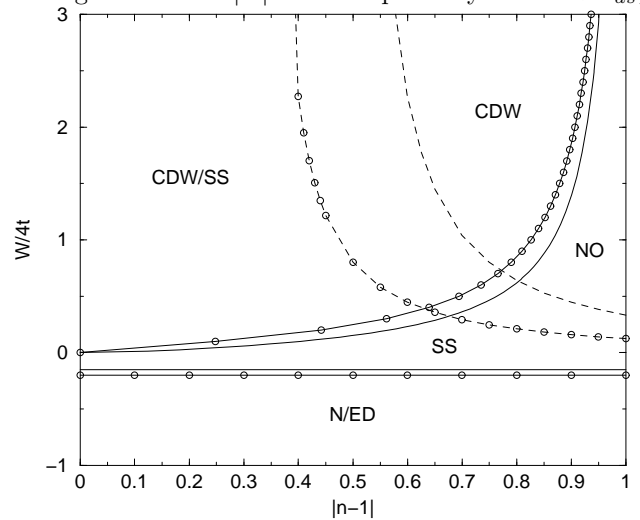}
\caption{\label{GPphDg} The phase diagram of the extended Hubbard
model solved in the mean-field approximation for $U/4t=-0.8$
(lines with no symbols) and $U/4t=-0.4$ (lines with circles). Full
lines are taken from the paper of Robaszkiewicz and
Pawlowski.~\cite{robaszkiewicz} Dashed lines come from
Eq.~(\ref{wcr}). N/ED denote phase separation between normal state
and electron droplets, SS - superconductivity, CDW - charge
density waves, CDW/SS - inhomogeneous phase separation between CDW
and SS, NO - normal state.}
\end{figure}

The third issue concerns the possible phase separation  for large,
 values of $|U|$ or $|W|$. This question was addressed by Hartree type analysis
e.g. in the paper of Robaszkiewicz and
Pawlowski~\cite{robaszkiewicz}, from where the phase boundaries
shown in Fig.~\ref{GPphDg} come (solid lines). Under the lowest,
horizontal line, for large n.n. attraction, we have a phase
separation of normal state and electron droplets. In the area
between the solid lines there is $s$-wave superconductivity
(on-site only, denoted as SS) and above the upper solid line there
is inhomogeneous phase separation of charge density waves (CDW)
and superconductivity (CDW/SS). For the half filled band the
ground state is CDW for $W>0$ while CDW and superconductivity are
degenerated for $W=0$.

Our second superconducting $s$-wave solution does not enter this
picture but for intermediate values of on-site attraction,
Eq.~(\ref{wcr}) helps us to make an upper boundary for
superconducting state, denoted by dashed lines in
Fig.~\ref{GPphDg}. Above this line no superconductivity exists.
This way for small electron densities, for $W$ large enough (above
the critical value) a normal state is obtained (denoted as NO in
the figure). With increasing $n$ we can have a transition to SS
and then to CDW/SS state. For certain larger $W$ we can have
NO-CDW/SS transition and for still larger $W$ we obtain
NO-CDW-CDW/SS transition. This way a phase diagram of
Ref.~\cite{robaszkiewicz} is substantially modified. Considering
full extended $s$-wave superconducting solution, with order
parameter depending on both on- and intersite interaction, changes
phase boundaries in Fig.~\ref{GPphDg} only in negligible way.

Let's note, that the described modification of the phase diagram
takes place only for intermediate values of $|U|$. For larger
values of $|U|$ or more precisely for $U<U_{as}$, where $U_{as}$
was defined in connection with Fig.~1, $s$-wave superconductivity
exists independently of $W$, and there is no normal phase area in
the phase diagram.

In conclusion a mean-field criterion for existence of $s$-wave
superconducting solution in the extended Hubbard model was
calculated. The modified $W-n$ phase diagram was shown. A second
$s$-wave solution was found and a threshold for its existence
together with a full phase diagram in the $UW$ plane were
calculated. The second solution helped to answer the question why
limits $U$ (or $W$) $=\pm\infty$ are formally the same. (These
results apply also to the two-electron bound states of the same
symmetry). First-order transition in the extended $s$-wave
solution was found and its possible implications on the crossover
problem were discussed. The results may apply to all properties of
the superconducting state of extended $s$-wave symmetry.

I acknowledge discussions with R. Micnas, S. Robaszkiewicz, P.
Grzybowski and B. Tobijaszewska-Chorowska and support from the
Foundation for Polish Science. This work was also supported by the
Polish State Committee for Scientific Research (KBN), Project No.
1 P03B 084 26.

\end{document}